\documentstyle[12pt]{article}
\newcommand{\beq}{\begin{equation}}
\newcommand{\eeq}{\end{equation}}

\newcommand{\GeV}{\,\mbox{GeV}}
\newcommand{\MeV}{\,\mbox{MeV}}
\newcommand{\matel}[3]{\langle #1|#2|#3\rangle}

\newcommand{\ra}{\rightarrow}

\begin{document}
\begin{titlepage}
\pagestyle{empty}
\begin{flushright}
\begin{tabular}{l}
CERN-TH.7082/93 \\
UND-HEP-93-BIG06\\
UMN-TH-1225/93\\
TPI--MINN--93/53--T\\
TECHNION-PH-93-40\\
\end{tabular}
\end{flushright}
\begin{center}
{\LARGE \bf The Baffling Semileptonic Branching Ratio of
$B$ Mesons}
\\
\vspace{0.6cm}

{\large Ikaros \ Bigi}\\
{\em TH Division, CERN,
CH-1211 Geneva 23, Switzerland
\footnote{During the academic year 1993/94}
}
\\
and \\
{\em Physics Dept.,
Univ. of Notre Dame,
Notre Dame, IN 46556, U.S.A. }
\footnote{Permanent address}
\\ {\em e-mail address: VXCERN::IBIGI, BIGI@UNDHEP}\\

\medskip

{\large Boris \ Blok}\\
{\em  Physics Department, Technion,
Haifa, Israel}
\\ {\em e-mail address:PHR34BB@TECHNION.BITNET}\\

\medskip

{\large Mikhail\ Shifman}\\
{\em Theor. Phys. Inst., Univ. of Minnesota,
Minneapolis, MN 55455, USA}
\\ {\em e-mail address: SHIFMAN@UMNACVX.BITNET}
\\

\medskip

{\large Arkady \ Vainshtein}\\
{\em Theor. Phys. Inst., Univ. of Minnesota,
Minneapolis, MN 55455, USA}\\
and\\
{\em Budker Inst. of Nucl.Physics, Novosibirsk 630090, Russia}
\\
{\em e-mail address:VAINSHTE@UMNACVX.BITNET}

\bigskip


{\bf Abstract}

\end{center}
{\small
\noindent
The apparent gap between the measured and the expected
value for the semileptonic branching ratio of $B$ mesons
has become more serious over the last year. This is due
to the improved quality of the data and to the increasing
maturity of the theoretical treatment of non-perturbative
corrections. We discuss various theoretical options
to reduce the semileptonic $B$ branching ratio; among
the more spectacular resolutions of the apparent puzzle is
the possibility of an unorthodox enhancement in
non-perturbative corrections or even of an intervention by
`New Physics'. Phenomenological implications of such scenarios
are pointed out.}


\vspace{0.5cm}
\noindent CERN-TH.7082/93  \\
October 1993

\end{titlepage}

\pagestyle{plain}
\setcounter{page}{1}
\section{The Problem}

Over the last few years
the measured semileptonic branching ratio of B mesons has
consistently turned out to be noticeably smaller than
theoretical expectations. Up until recently this could
be waved off as no worse than an embarrassment for theory
or experiment since both were somewhat uncertain in their
pronouncements. Now the
situation has changed in two respects: on the one hand
the data became more
mature, both statistically and systematically;
on the other hand a theoretical machinery has been developed
that is genuinely based on QCD and that
allows treating non-perturbative
corrections to inclusive heavy flavour decays in a
quantitative and systematic way \cite{BIGI,BLOK,BIGI1}.

The situation is as follows:
A `model-independent' ARGUS analysis yields \cite{ARGUS}
$$BR_{SL}(B)= 9.6\pm 0.5\pm 0.4\% \eqno(1)$$
whereas the CLEO collaboration finds \cite{CLEO}
$$BR_{SL}(B)= 10.65\pm 0.05\pm 0.33\% \eqno(2)$$
using the model of Altarelli {\em et al.}
\cite{ACM} for the $shape$ of the
lepton spectrum. One should keep in mind
that this model provides a good
approximation to the true QCD lepton spectrum as
calculated through a $1/m_Q$ expansion \cite{SPECTRUM}.
The present data thus clearly suggest:
 $$BR_{SL}(B)|_{exp} \leq 11\% .\eqno(3)$$

In a naive parton model  where even
perturbative QCD is ignored one obtains
$$BR(b\rightarrow cl\nu)\simeq 15\div 16\% ,\eqno(4)$$
i.e. a non-leptonic enhancement of $\sim 50\%$ has
to be found to reproduce the data.

The main assertions of this paper are:

$\bullet$ Non-perturbative corrections
affect inclusive non-leptonic widths of $B$ mesons
only on the few per cent level.
To first approximation they can  be ignored in
calculating $BR_{SL}(B)$.
They cannot reduce the prediction
to the 11\% level or below -- as long
as QCD can be treated in a `standard' fashion to be
defined later.

$\bullet$ It is then mainly
the {\em perturbative} corrections that
control the size of $BR_{SL}(B)$. They indeed generate a
non-leptonic enhancement thus
reducing $BR_{SL}(B)$.
At present there are still some missing pieces
in the perturbative
analysis; yet making reasonable conjectures
about them one can conclude
$$BR_{SL}(B)|_{QCD} \geq 12.5\% .\eqno(5)$$

$\bullet$ An intriguing problem has
arisen, which warrants serious consideration: how can one
find an additional non-leptonic
enhancement of at least 15 to 20\% to satisfy
the bound of eq. (3)?

$\bullet$ A priori an explanation could invoke
one of two major surprises, namely the existence of
`anomalously' large non-perturbative contributions from
QCD -- the more conservative of the two options --
or the intervention of some new interactions coupling only
to quarks, but not to leptons -- clearly the more radical
option.

$\bullet$ Neither of these options appears particularly
natural. Since they are supposed to generate
at least $\sim$ 20\% of all $B$ decays
they could well lead to further
phenomenological consequences: lifetimes differences
between $B^-$ and $B_d$ mesons of 20-30\% rather than
the expected < 10\%; likewise lifetime differences
between $\Lambda _b$ and $B_d$ that exceed 10-15\%.
The features of non-leptonic final states -- say the
charm content or decay multiplicities -- should exhibit
some significant differences to what is expected in the standard scenario.

The remainder of this paper will be organized as follows:
in Sect. 2 we discuss the perturbative
corrections;
in Sect. 3 we analyse the size of various non-perturbative
corrections;  in Sect. 4 we describe phenomenological
consequences of various possible resolutions for the
puzzle posed by the observed semileptonic branching ratio
before giving an outlook in Sect. 5.

\section{General Procedure and The Leading
Perturbative Corrections to $BR_{SL}(B)$}

The transition {\em operator}
$\hat T(b\ra f\rightarrow b)$ describing
the forward scattering of $b$ quarks via an intermediate state
$f$ to second order in the weak interactions is given by
\cite{ORIGIN}
$$\hat T(b\ra f\rightarrow b)
= i\int d^4x \{ {\cal L}(x) {\cal L}(0)\}_T\eqno(6)$$
with ${\cal L}$ denoting the relevant effective weak
Lagrangian and $\{.\}_T$ the time-ordered product.
A Wilson operator expansion (OPE) allows the expression of the
non-local operator $\hat T$ as the infinite sum of local
operators of increasing dimension with
coefficients that contain
higher and higher
powers of $1/m_b$. Long distance dynamics determines the
on-shell matrix elements of these local operators whereas
short distance dynamics controls their c number coefficients.
One conventionally computes the latter in perturbative
QCD; we refer to this procedure as the `standard'
prescription for QCD. It is by no means exact: there
are, even at short distances, non-perturbative contributions
that affect the coefficient functions. They are however
estimated to be of no practical significance in $B$
decays -- a point to which we will return later on.

The lowest dimensional operator that appears in the OPE and
dominates for $m_b\rightarrow \infty$
is $\bar bb$. Flavour symmetry
fixes the leading term in its matrix element:
$$\langle B|\bar bb|B\rangle /(2M_B)
=1+ {\cal O}(1/m_b^2)\; ,\eqno(7)$$
where we have used the relativistic normalization for
the $B$ meson state.
It is this term that reproduces the Spectator Model;
the coefficient of $\bar bb$ thus represents the purely
perturbative corrections.

The most detailed perturbative analysis of $BR_{SL}(B)$ in
the parton model has been undertaken in ref. \cite{AP}
(AP in what follows).
We will critically review its main points.

{}From the Lagrangian for semileptonic $b\ra c$ transitions
$${\cal L}_{SL} = \frac{G_FV_{bc}}{\sqrt{2}}
(\bar c\Gamma_\mu b)(\bar e\Gamma_\mu \nu )\; \eqno(8)$$
one obtains the semileptonic width of $B$ mesons:
$$
\Gamma_{SL}=\Gamma(b\rightarrow c l{\bar\nu}_l) =
\Gamma_0I_0\left ( \frac{m_c^2}{m_b^2} ,\frac{m_l^2}{m_b^2}, 0
\right )
\left [ 1-\frac{2\alpha_s}{3\pi}\right .
f\left (\frac{m_c^2}{m_b^2},\frac{m_l^2}{m_b^2}\right )
+\left .{\cal O}(\alpha_s^2)\right ],
\eqno(9)$$
where we have used a notation similar to that of AP:
$$
\Gamma_0\equiv \frac{G_F^2m_b^5|V_{bc}|^2}{192\pi^3}\; \; ;
\eqno(10)$$
the phase-space factor $I_0$ accounts for the masses of the
fermions in the final state \cite{PHAM}.
The subscript 0 indicates that $I_0$ is the
phase-space factor in the `parton'  expression for $\Gamma$.
In the electronic and muonic semileptonic decay
rates we can neglect the lepton masses; this leads to the simple
expression
$$
I_0(x,0,0) = (1-x^2)(1-8x+x^2)-12x^2\ln x.
\eqno(11)$$
With $\tau$ leptons in the final state we need to know $I_0(x,y,0)$;
its explicit expression can be found in ref. \cite{PHAM}.
The function $f$ plays the analogous role in the ${\cal O}
(\alpha_s)$ term,
$$
f(0,0) =\pi^2-\frac{25}{4}.
$$
There are two classes of non-leptonic decays. The effective
weak Lagrangian for $b\ra c\bar ud$ transitions is
given by
$${\cal L}(\mu) = \frac{G_F}{\sqrt{2}}V_{bc}V_{ud}(c_1 {\cal O}_1 +
c_2 {\cal O}_2)
\eqno(12)$$
where ${\cal O}_{1,2}$ are operators,
$$
{\cal O}_1 = (\bar c\Gamma_\mu b)(\bar d\Gamma_\mu u ) , \,\,\,
{\cal O}_2 = (\bar{c}_i\Gamma_\mu  b^j)(\bar{d}_j\Gamma_\mu
u^i).
\eqno(13)$$
with $\Gamma_\mu = \gamma_\mu (1+\gamma_5)$.
The Wilson coefficients $c_{1,2}$ account for the
radiative corrections from virtual gluon momenta
from $\mu$ up to $M_W$; they have been determined
from perturbation theory
\cite{AM,BURAS}:
$$c_1 =\frac{1}{2}(c_++c_-),\;
c_2 =\frac{1}{2}(c_+-c_-), \;
c_{\pm} = \left[
\frac{\alpha_s (\mu )}{\alpha_s (M_W)}\right]^{d_{\pm}}.
\eqno(14)$$
(The penguin contribution showing up at the 1\% level is omitted).
The non-leptonic enhancement factor (beyond the global colour
factor $N_C$) is then given by
$$
\eta = \frac{c_-^2+2c_+^2}{3}. \eqno(15)$$

The $b\ra c\bar cs$ transitions are treated in a
completely analogous fashion with the obvious
substitutions of $\bar c$ for $\bar u$ and
$s$ for $d$.

For the non-leptonic widths one then obtains:
$$\Gamma(b\rightarrow c\bar u d) + \Gamma(b\rightarrow c\bar u s)
=3\Gamma_0 \, I_0\left ( \frac{m_c^2}{m_b^2}, 0,0
\right ) \, \eta \, J.
\eqno(16)$$
For the channel $b\ra c\bar cs$
an analogous expression holds, with the substitution:
$$
I_0\left ( \frac{m_c^2}{m_b^2}, 0,0\right )
\ra I_0\left ( \frac{m_c^2}{m_b^2},\frac{m_c^2}{m_b^2},0
\right ) ,
\eqno(17)$$
where
$$
I_0(x,x,0) =v (1-14x-2x^2-12x^3)+24x^2(1-x^2)\ln \frac{1+v}{1-v},
\eqno(18)$$
$$
v=\sqrt{1-4x}.
$$
A few remarks are in order concerning eqs. (16,17):

\vspace{0.3cm}

\noindent (i) In the phase-space
factor $I_0$ the light quark masses are neglected.
To obtain a self-consistent QCD treatment one has to employ
current quark masses;
since $m_s^2/m_b^2\sim 10^{-3}$ one can
then ignore even the strange
quark mass.

\noindent (ii) The enhancement factor $\eta$
is produced by the anomalous dimensions of the operators
in the effective weak Lagrangian ${\cal L}(\mu )$ (see eq. (12))
in the leading log approximation,
with $\mu$ the normalization point.

\noindent
(iii) The last factor, $J$, represents the next-to-leading
corrections. These appear as $\alpha_s$ contributions
in the effective Lagrangian ${\cal L}$
(coming, in particular, from next-to-leading
terms in the anomalous dimensions), as well as $\alpha_s$
corrections in the calculation of the non-leptonic width $\Gamma$,
eq. (16).
For massless quarks in the final state the expression for
$J$ simplifies considerably and is given in AP.
We are using this expression for $b\ra c\bar ud$ as well
as for $b\ra c\bar cs$ transitions.

The effective Lagrangian ${\cal L}(\mu)$
includes effects due to gluon exchanges with virtual momenta
from $M_W$ to $\mu$;  loop momenta below $\mu$ should be taken into
account in the evaluation of $\Gamma$. The physical result,
the product $\eta J$, must not depend on $\mu$, of course, and it is
the $\mu$ dependence of the factor $J$ that compensates for the $\mu$
dependence of $\eta$, eq. (14).

The concrete expressions for $\eta$ and $J$ derived and used in AP
satisfy the property of $\mu$ independence of $\eta J$
to order $\alpha_s$, but not $\alpha_s^2$. This is the reason why the
non-leptonic widths obtained in AP depend on the choice of $\mu$,
the variation of $\eta J$ being rather significant numerically.
It is quite conceivable that there is a single value of $\mu$
which, when
substituted in $\eta J$,  reproduces the
correct coefficient in
the $\alpha_s^2$ terms in $\eta J$.
Since the $\alpha_s^2$ terms are unknown
at the moment, one can only speculate on what this value of $\mu$
might be,
relying on heuristic arguments.

For years it was assumed that the appropriate choice is $\mu = m_b$.
If one then uses the anomalous dimensions obtained in the leading
\cite{AM} and next-to-leading  \cite{BURAS}
approximations one arrives at
$$
\eta \approx 1.1,\,\,\, J\approx 1.15 .
\eqno(19)$$
(Notice that the next-to-leading order effect is stronger than the
leading one; yet both are relatively small.)

The aim of the authors of ref. \cite{AP} was to push
the theory
to the extreme values it can produce.
To this end they have chosen the
normalization point $\mu$  as  low as $m_b/2$.
Then in the scenario with
$\alpha_s (M_Z) = 0.125$ -- which is somewhat on the
large side of the present world average --
the enhancement factors $\eta$
and $J$ are both increased:
$\eta\approx  1.27$, $J\approx 1.19$. The difference
between $1.1\times 1.15 = 1.26$ and $1.27\times 1.19 = 1.51$  measures the
uncertainty in the coefficient of $(\alpha_s/\pi)^2$. Notice that the
two options considered in AP correspond to the difference of roughly 30
in this coefficient!

It might be tempting to motivate the choice $\mu =m_b/2$
as follows. In at least
a part of the next-to-leading corrections the characteristic off-shellness
is smaller than $m_b^2$. Consider, for instance, the diagram of fig. 1,
where the gluon is exchanged between the $u$ and $d$ lines. (Let us
note in passing that for the {\em one}-gluon exchange this is the
only correction contributing to the ratio
$\Gamma (b\rightarrow c\bar u d)/\Gamma_{SL}$.) This correction is
identical to the ${\cal O}(\alpha_s)$ term
in the ratio $R=
\sigma (e^+e^-\rightarrow {\rm hadrons})/\sigma (e^+e^-
\rightarrow\mu^+\mu^-)$ and is equal to
$$
 1+\frac{\alpha_s}{\pi} +{\cal O}(\alpha_s^2).
\eqno(20)$$
The $\mu$ dependence of $\alpha_s$ is hidden in the $\alpha_s^2$
terms. In $e^+e^-$ it is known that choosing the argument of
$\alpha_s$ to be equal to the
invariant mass of the quark pair does not lead to large
$\alpha_s^2$ terms. In $b$ decays the invariant mass
of the $\bar u d$ pair is integrated over a range limited
from above by $m_b$. A characteristic value of the invariant mass
is close to $m_b/2$.

The effective reduction of $\mu$ does not apply, however, to other
contributions. An example of a graph with a typical virtual mass of
$m_b^2$ is shown in fig. 2, where the closed circles denote
the four-fermion vertices from the effective weak
Lagrangian, eq. (12). In this
diagram the contribution from gluons
with loop momenta below
$m_b$ is suppressed in a power-like way.

Setting $\mu =m_b/2$ in the whole expression for $\eta J$
thus represents an unnatural or `twisted arms' scenario.
We conclude that this
`twisted arms scenario' most likely yields an
{\em over}estimate for the enhancement
factor $\eta J$. Non-extreme estimates presented in AP, with a lower
value of $\alpha_s (M_Z)$ and the normalization point at $m_b$, result in a
weaker enhancement of the non-leptonic channels corresponding to
$$BR_{SL}(B) \geq 12.5\% \eqno(21)$$
with the lower bound attained for
$$m_b\simeq 4.6\; \GeV ,\; \; \; m_c\simeq 1.2\; \GeV ,\; \; \;
m_s\simeq 0.15\; \GeV ,\; \; \; m_{u,d}\simeq 0\; .\eqno(22)$$
Unless one has actually computed the $\alpha_s^2$ terms,
one cannot make
categoric statements; nevertheless it seems fair to say
that the natural  prediction
of perturbative QCD for $BR_{SL}(B)$ exceeds the experimental number
by at least 1.5 percentage points.

It would seem natural
-- and up until recently it would have been quite
appropriate -- to
attribute the remaining difference between
the expectation expressed in eq. (21) and the data,
eq. (3), to non-perturbative corrections
further enhancing the non-leptonic width.
In the next section we will show that such a
`deus ex machina' is unlikely to work this
time around, at least not in `standard' QCD.
This opens a window for exotic mechanisms which
might
contribute as much as 20 to 30\% of the total non-semileptonic width.
Below it will be argued that `standard' non-perturbative
effects cannot explain such a large gap.

\section{Non-perturbative Corrections to $BR_{SL}(B)$}

As stated in the previous section, non-perturbative corrections
due to soft quark-gluon interactions are incorporated through
the appearance of higher-dimensional local operators in the
OPE and through the $B$ meson expectation values of all
operators, including $\bar bb$. We have already mentioned
that the c-number coefficients in the OPE are computed
perturbatively and that we refer to this prescription as
the `standard' version of QCD \cite{SVZ}.

Since there is no dimension four operator that can
contribute to $\hat T(b\ra f\ra b)$
\cite{TM,BIGI,BLOK},
non-perturbative corrections
to totally integrated rates
appear first at the $1/m_b^2$ level through the matrix
elements of dimension five operators.
The absence of corrections of order $1/m_b$ has two important
consequences: (i) The natural scale for non-perturbative
corrections in beauty decays is of order a few per cent:
$(\mu _{had}/m_b)^2\sim 0.04$ for $\mu _{had}\sim 1\; \GeV$.
(ii) It establishes that one has to use current quark masses
for a self-consistent QCD treatment and thus removes a
conceptual ambiguity inherent in phenomenological models.

The $1/m_b^2$ corrections have already
been analysed in the literature and we will review them
here. In addition we will estimate the
contributions from dimension six operators.

The semileptonic and non-leptonic
widths through order $1/m_b^2$ are given by:
$$\Gamma _{SL}(B)=\Gamma _0
\cdot I_0(x,0,0)\frac{\matel {B}{[\bar bb]_{\Sigma}}{B}}{2M_B}
\eqno(23a)$$
$$\Gamma _{NL}(B\ra [C=1])=
\Gamma _0\cdot N_C\cdot
\left\{ A_0I_0\frac{\matel {B}{[\bar bb]_{\Sigma}}{B}}{2M_B} -
\frac {8A_2I_2(x,0,0)}{m_b^2}
\frac{\matel{B}{{\cal O}_G}{B}}{2M_B}\right\} ,
\eqno(23b)$$
$$[\bar bb]_{\Sigma}=\bar bb-
\frac{2[I_0-\frac{1}{2}x\frac{d}{dx}I_0]}{I_0}
\cdot
\frac{{\cal O}_G}{m_b^2},\; \;
{\cal O}_G=\frac {1}{2}\bar bi\sigma \cdot Gb\eqno(23c)$$
where $I_0$ and $I_2$ are phase-space factors:
$I_0(x,0,0)$ is defined in eq. (11) and
$$I_2(x,0,0)=(1-x)^3,\,\,\, x=(m_c/m_b)^2;$$
$A_0=\eta J$ and
$A_2=(c_+^2-c_-^2)/2N_C$ represent the radiative QCD corrections.
Due to the colour flow, the operator ${\cal O}_G$ in
eq. (23b) arises from the interference of the two operators
$O_1$ and $O_2$, eq. (13), see refs. \cite{BIGI,BLOK}.

The matrix element $\matel{B}{[\bar bb]_{\Sigma}}{B}$
enters as an overall factor into both the semileptonic
and non-leptonic width; its value does therefore not
affect the branching ratio.
Furthermore $\matel {B}{{\cal O}_G}{B}$ can be determined from
the observed $B^*-B$ mass splitting since ${\cal O}_G$
represents the chromomagnetic operator
(${\cal O}_G\ra -\bar b\vec \sigma \cdot \vec Bb$ in the
non-relativistic limit):
$$
\frac{1}{2M_B}<B|{\cal O}_G|B> \equiv \mu_G^2 =
\frac{3}{4}(M_{B^*}^2-M_B^2)
\simeq 0.37\,\, {\rm GeV}^2 .
\eqno(24)
$$
Altogether one thus finds through order $1/m_b^2$
$$\Gamma _{SL}(B)\simeq \Gamma _0
\cdot \frac{\matel{B}{\bar bb}{B}}{2M_B}\cdot
[I_0(x,0,0)+
\frac{\mu _G^2}{m_b^2}(x\frac{d}{dx}-2)I_0(x,0,0)]
\eqno(25a)$$
$$\Gamma _{NL}(B)\simeq \Gamma _0
\cdot N_C\cdot \frac{\matel{B}{\bar bb}{B}}{2M_B}
\cdot \{ A_0[\Sigma I_0(x)+\frac{\mu _G^2}{m_b^2}
(x\frac{d}{dx}-2)\Sigma I_0(x)]-
$$
$$
8A_2\frac{\mu _G^2}{m_b^2}\cdot
[I_2(x,0,0)+I_2(x,x,0)]\} \eqno(25b)$$
with $\Sigma I_0(x)\equiv I_0(x,x,0)+I_0(x,0,0)$,
see eqs. (11,18). The contributions from
$b\ra c\bar cs$ transitions are included through
$I_0(x,x,0)$ and
$$I_2(x,x,0)=v\left ( 1+\frac {x}{2}+3x^2\right )
-3x(1-2x^2)\log
\frac {1+v}{1-v} \; .\eqno(26)$$
It is evident from eq. (25b) that the operator ${\cal O}_G$
generates a non-leptonic enhancement since $A_2<0$. We will
now discuss how large such an effect could be, with a bias
towards enhancing this correction as much as
reasonably possible. This bias expresses itself in the choice of the
scale $\mu$ and the values for $m_b$ and $m_c$.

Following ref. \cite{V} we adopt
$$ m_b^{(pole)}=4.8\; \GeV . \eqno(27)$$
{}From the observed $B-D$ mass difference one deduces
$m_b^{(pole)}-m_c^{(pole)}\simeq 3.34\; \GeV$ and thus
$m_c^{(pole)}\simeq 1.45\; \GeV$.
The choice of the pole mass for charm is not quite appropriate
for $B\ra D+X$ decays
since the effective  off-shellness of charm quark
is of order $m_b^2/2$. We will therefore use
$$m_c^{(eff)}\sim 1.35\; \GeV .\eqno(28)$$
Such values for $m_c$ lead to quite a sizeable weight for
$b\ra c\bar cs$ transitions, namely close to one half of that
for $b\ra c\bar u d$
(although it is a little bit smaller than in AP);
we will return to this point
later on.

Adopting a scale $\mu$ as low as $m_b/2$ (and $\alpha_s (M_Z) =0.125)$
in the leading-log
expression for $c_{\pm}$ we get
$$c_+\simeq 0.85, \; \; \; \; c_-\simeq 1.45\; .
\eqno(29)$$
Putting everything together we find
$$\delta BR_{SL}(B)\sim -0.02 BR_{SL}(B)\sim - 0.003 , \eqno(30)$$
i.e. the leading non-perturbative correction cannot close the gap
between the theoretical expectation and the present trend in
the data.

One then turns to discussing non-perturbative
corrections induced by higher-dimensional operators. There one has
to analyse anew only those contributions
that are
non-factorizable,
i.e. where
the $\bar u d$ quark loop is connected to the rest of the diagram.
Those corrections that are localized `inside' this loop -- and these
factorizable non-perturbative corrections certainly do exist -- are the
same as in $e^+e^-$ annihilation cross sections or $\tau$ decays. In the
integrated rate they are known \cite{PICH}
not to exceed $\sim 2$\% and can be
disregarded. Let us also note in passing that the factorizable
condensate
corrections start from $m_Q^{-4}$ \cite{SVZ}, and that the hard
non-perturbative effects
are suppressed by even higher powers of $m_Q^{-1}$ \cite{NSVVZ}.

There are two classes of dimension six operators producing
$1/m_b^3$ corrections, namely

$\bullet$ four-quark operators
$${\cal O}_{4q}=(\bar b\Gamma q)(\bar q\Gamma b), \eqno(31)$$
with $q$ denoting light-quark fields and $\Gamma$ a combination
of $\gamma$  and colour matrices; they are generated by
one-loop graphs as shown in fig. 3.

$\bullet$ Quark-gluon operators containing $\bar b$ and
$b$ fields, the gluon field strength tensor $G_{\mu \nu}$
and an additional covariant derivative. These operators arise
from two-loop diagrams, as shown in fig. 4; hence their coefficients
are numerically quite suppressed relative to those of the
four-quark operators. Using the equations of motion,
in particular
$$
iD_0 b =-\frac{(\vec\sigma\vec D )^2}{2m_b} b +{\cal O}(m_b^{-2}) ,
\eqno(32)$$
it can be shown \cite{BS} there are only two
spin-zero quark-gluon operators of dimension six, namely
$$
\bar b (D_\mu G_{\mu\nu})\Gamma_\nu b
\eqno(33)$$
and
$$
{\cal O}_E = \bar b\sigma_{\mu\nu}G_{\mu\rho}\gamma_\rho
iD_\nu b
\rightarrow \bar b\vec\sigma \, \vec E \times i \vec D b ,
\eqno(34)$$
where $\vec E$ is the chromoelectric field. All other
dimension six quark-gluon operators
can be shown to be reducible to the operators listed above.
The operator $\bar b (D_\mu G_{\mu\nu})\Gamma_\nu b$ is actually
a four-quark operator since
$$
D_\mu G_{\mu\nu} =  - g^2\sum \bar q \gamma_\nu T^a q .
\eqno(35)$$
Since its coefficient contains an extra factor of
$\alpha_s/\pi$, compared with
the four-quark operators coming from the one-loop graphs,
its contribution can be ignored.

(i) To evaluate $\Delta\Gamma_{4q}$,
the contributions of the four-quark operators to
the width,
we use factorization, or the vacuum saturation approximation,
$$
\matel{B}{(\bar b\Gamma q )(\bar q\Gamma b)}{B} \approx
\matel{B}{(\bar b\Gamma q )}{0}\cdot
\matel{0}{(\bar q\Gamma b)}{B}.
\eqno(36)$$
In this approximation those four-quark operators
that represent the `Weak Annihilation' mechanism
give a very small
contribution, which is also helicity-suppressed by
$m_c^2/m_b^2$ \cite{URI}. Such four-quark
operators will be disregarded. The ones that survive
are due to the
interference mechanism, see fig. 3b. There are actually
two such operators differing in their colour flow with
Wilson coefficients $K_1$ and $K_2$. Their expressions
are given in refs. \cite{GRT,VS} with the
normalization point chosen at $m_b$:
$$K_1=\frac{1}{3}(2c_+^2-c_-^2), \; \;
K_2=c_+^2+c_-^2.$$
The matrix element of the four-quark operators is
expressed as follows:
$$
\mu_{4q}^3 =\frac{1}{2M_B}\matel{B}{\bar b
\Gamma_\mu u \bar u \Gamma_\mu b}{B}\approx \frac{1}{2}f_B^2M_B .
\eqno(37)$$
It should be emphasized that the four-quark contribution of this type
exists only for $B^-$ and is absent for $B^0$ mesons.
There is a technical subtlety involved in making the
factorization ansatz: matrix elements have an implicit
dependence on the normalization scale. As far as the
strong interactions are concerned, $m_b$ is a completely
foreign parameter. It is much more natural to adopt
eq. (37) at a typical hadronic scale $\mu$. The four-fermion
operators have then to be evolved down to $\mu$. This
is achieved by hybrid renormalization \cite{HYBRID} computed
in the leading-log approximation \cite{VS};
its effects get included in the quantities $K_1$
and $K_2$. The
inclusion of this hybrid renormalization turns out to
be numerically relevant, too: for they remove an
accidental cancellation in the strength of the
destructive interference.

So far the quantity
$f_B$ has not been measured yet. Its value is estimated
via QCD sum
rules and via QCD simulations
on the lattice. The recent and most reliable estimates
cluster around 190 MeV for QCD sum rules \cite{NEUBERT}
and in lattice calculations \cite{BLS}. Taking this
interval to represent the measure of uncertainty we get
$$
\mu_{4q}^3 \sim 0.1 \,\, {\rm GeV}^3 .
\eqno(38)$$
and thus
$$\frac{\Delta \Gamma_{4q}(B^-)}{\Gamma (B^-)}
\simeq - 0.05\cdot (f_B/200\; \MeV)^2\eqno(39)$$
$$
\Delta \Gamma_{4q}(B^0)\approx 0 . \eqno(40)
$$
One should notice that this correction {\em suppresses} the $B^-$
non-leptonic width. Therefore it
works in a direction opposite to the `desired' one -- enhancement of
the
non-leptonic width!

The correction in the
non-leptonic width due to the four-quark operators of
dimension six are not smaller than those due to the
dimension five operator ${\cal O}_G$, see eq. (29).
This can be understood in the following way:
the Wilson coefficient $c_{4q}$ is determined by
one-loop graphs while $c_G$ is extracted from a two-loop diagram.
For the higher dimensional operators the hierarchy of
corrections is expected to be normal:
terms of higher order in $1/m_b$ are numerically
smaller.

The four-quark operator considered above is the first to
differentiate between
$B^-$ and $B^0$ lifetimes.  The estimate
given above is quite consistent with
recent data \cite{BESSON} for the ratio of the lifetimes,
 $ \tau (B^0)/\tau (B^-) =1.05\pm 0.16\pm 0.15$.

We can estimate the matrix element
$\mu_E^3 =\matel{B}{{\cal O}_E}{B}/2M_B$ -- where
${\cal O}_E$ denotes the dimension six
quark-gluon operator -- in two complementary ways:
treating the light quarks in the $B$ meson in the
relativistic limit one finds that
the chromo-electric field in the light cloud is of the same order
as the chromo-magnetic field. If so, $\mu_E^3$ differs from $\mu_G^2$
(see eqs. (24,34)) by the average quark momentum $\mu_\pi$,
$$
\mu_E^3\sim \mu_G^2 \mu_\pi .\eqno(41)
$$
It is not difficult to check that the opposite limit of
non-relativistic {\em light} quarks leads to the same estimate.

Accordingly one concludes
$$
\frac{\Delta\Gamma_E}{\Delta\Gamma_G}\sim \frac{\mu_\pi}{m_b}
\sim 0.1, \eqno(42)
$$
a small correction to a correction in the non-leptonic width,
which by itself is about 3\%.
Notice that $1/m_b$ terms in the matrix element of ${\cal O}_G$
are of the same order.

{\em Let us summarize the discussion of the last two sections}:
using our best theoretical judgement we conclude that
$BR_{SL}(B)$ is expected to exceed 12\%. In the present data
$BR_{SL}(B)$ is seen to fall below 11\%.
There are several possible scenarios for
closing the gap between expectation and observation:

(i) Improved data could move $BR_{SL}(B)$ above 12\%.

(ii) The width for $b\ra c\bar cs$ transitions is
larger than anticipated due
to larger than expected non-perturbative
corrections in that channel. (One should keep in mind that
our treatment of non-perturbative corrections, which is
based on a large energy release in the decay, is somewhat
less reliable in $b\ra c\bar cs$.) If such an
enhancement of $\Gamma (b\ra c\bar cs)$ were the cause
of the puzzle, it would lead to an obvious
consequence: it would considerably {\em enhance} the
charm multiplicity over what is expected --
and that already comes out too large
compared to what is observed, see below!

(iii) Instead of a single-source resolution of the
apparent puzzle, there could be a `cocktail', i.e.
a combination of several small effects all working in the
same direction: the experimental number could inch up;
higher order non-perturbative corrections
could turn out to be abnormally large and all positive in
$\Gamma_{NL}$; last, but not least, next-to-leading
perturbative corrections could be more sizeable than anticipated.
Note, however, that each ingredient of the
`cocktail' affects the branching ratio at the level $\sim$ 0.1 to 0.2\%.

(iv) Non-perturbative corrections
could be {\em dramatically} larger than anticipated. This certainly would
require going beyond the standard version of OPE. As mentioned
before, in general there are non-perturbative short-distance
contributions to the Wilson coefficients (which are sometimes
referrred to as `hard' non-perturbative terms \cite{NSVVZ}).
The hard non-perturbative
terms can show up in the coefficient functions of the operators
$\bar b b$, $\bar b \sigma G b$ and/or $\bar b q \bar q b $. In the
latter two cases they must enhance the coefficients by a factor of $\sim$ 5
(and change the sign of the four-fermion coefficient)
to ensure the 20\% enhancement of the non-leptonic widths.
To attribute $\sim 20\;\%$ non-leptonic enhancement to such
`non-standard' terms would be very surprising, since
 they represent at most a 2-3\% effect in $\tau$ decays and
should even be more suppressed at the higher mass scale of
$B$ decays.

It is true that in $\tau$ decays quark-antiquark states
necessarily emerge as a colour-singlet whereas in $B$
decays also colour-octet configurations are possible.
It would, however, seem quite contrived to attribute an effect of the
alleged magnitude to this distinction. Yet if such an
unorthodox and unforeseen feature of QCD were responsible
for an additional non-leptonic enhancement, then
it should generate lifetime differences between
$B_d$ and $B^-$ mesons and/or between mesons and baryons at
the level of 15 to 30\% (if the non-perturbative hard terms enhance
$\bar b\sigma G b$ or $\bar b q\bar q b $).

(v) The most intriguing  possibility
would be the intervention of New Physics in B decays. This might
lead to a different charm content in the final state.

In the next section we will address the phenomenological
implications of these scenarios in some more detail.

\section{Phenomenological Implications}

We will discuss here three phenomenological aspects
of beauty decays, namely

$\bullet$ the charm content of the final state in
$B$ decays;

$\bullet$ charmless two-body decays of $B$ mesons;

$\bullet$ lifetime ratios, in particular
$\tau (B^-)$ vs. $\tau (B_d)$ and
$\tau (\Lambda _b)$ vs. $\tau (B_d)$.

\noindent (i) Lowering $m_c$ relative to $m_b$ will enhance
the weight of the non-leptonic $b\ra c\bar cs$
transition and thus reduce the expected semileptonic
branching ratio. By the same token it will enhance
considerably the charm content in the final state:
for the values of $m_c/m_b$ adopted in eqs. (27,28) one finds
$$ N_{charm}\equiv
\frac {{\rm Number\; \; of\; \; charm\; \; states}}{B\;
{\rm decays}}
\sim 1.2 \div 1.3 \eqno(43)$$
The data exhibit a considerably lower charm content, namely
\cite{Danilov}
$$ N_{charm}=0.932 \pm 0.10\; \; \; \; {\rm ARGUS}\eqno(44a)$$
$$ N_{charm}=1.026 \pm 0.057\; \; \; \; {\rm CLEO}\eqno(44b)$$
One should keep in mind that there are still
considerable uncertainties in the {\em absolute} value
of the charm branching ratio, in particular for
$D_s$ and $\Lambda _c$ decays. The errors quoted
above could well be underestimated. Yet even so, there
is no sign of an over-abundance of charm states in $B$
decays -- on the contrary there is some evidence for a
serious `charm deficit'! It is quite tempting to take this
as indirect evidence for the rather massive intervention
of New Physics. If the putative New Physics is
postulated to couple only to {\em non-charm quarks},
but not to leptons, and to provide $\sim 20\%$ of the
total decay rate, then $BR_{SL}(B)$ is lowered by
$\sim 20\%$, of course; yet at the same time the
charm deficit has evaporated. On the other hand
there is a certain constraint on such an
exciting scenario; this will be discussed next.

\noindent (ii) Strong penguin transitions of the type $b\ra s+g$
would seem to fit the bill: they contribute predominantly
to non-leptonic decays without charm states. In the Standard
Model one estimates them to contribute not more than 1\%
of the total width. In principle there could be New Physics
entering the internal loops inducing a penguin operator driving
20\% of all $B$ decays. Yet if that is the case,
one should wonder about the impact of such an enhanced
operator on the exclusive channel $B\ra K \pi$.
CLEO \cite{TWOBODY} has found
evidence for $B\ra K\pi \; +\; \pi \pi$ coupled with
upper bounds on the individual channels:
$$BR(B_d\ra \pi ^+\pi ^- + K^+\pi ^-)=
(2.4\pm 0.7\pm 0.2)\cdot
10^{-5}\eqno(45a)$$
$$BR(B_d\ra \pi ^+\pi ^-)\leq 2.9\cdot 10^{-5}\eqno(45b)$$
$$BR(B_d\ra K^+ \pi ^-)\leq 2.6\cdot 10^{-5}\eqno(45c)$$
$$BR(B_d\ra K^+ K^-)\leq 0.7\cdot 10^{-5}\eqno(45d)$$
These numbers are quite consistent with Standard Model
expectations, which, however, suffer from sizeable
uncertainties. Nevertheless a
`Scylla and Charybdis'
conundrum has to be a concern for all New Physics scenarios:
if New Physics prefers to couple to non-charm states in the
inclusive rate, where is its impact
on the exclusive two-body modes
$B\ra K\pi ,\,\pi \pi $?

\noindent
(iii) Rather smallish lifetime differences have been predicted
among beauty hadrons: $\tau (B^-)/\tau (B_d)\simeq
1+ 0.05\cdot (f_B/200\; \MeV)^2$ and
$\tau (\Lambda _b)/\tau (B_d)\sim 0.85-0.9$. If on the other
hand QCD contains some unforeseen non-perturbative features
that can lower the semileptonic branching ratios by
$\sim 20\;\%$, those could impose the lifetime
differences of 15 to 30\%.

\section{Summary and Outlook}

For several years the observed value for $BR_{SL}(B)$ has
been below the theoretically expected one. We think
that the data and the relevant theory have reached such a level of
maturity such that the apparent 20\% or so gap between
$BR_{SL}(B)|_{exp}$ and $BR_{SL}(B)|_{QCD}$ -- while
not absolutely conclusive yet -- has to be perceived as a serious
problem. If improved data do not move to higher values,
there are three possible resolutions of such a discrepancy:

(i) `The dull way out':
Several effects -- each of order a few per cent --
`cooperate' to generate a 20\% correction.
There would be no other interesting/clear
phenomenological implication.

(ii) `The tantalizing resolution':
Corrections due to higher
dimensional operators and/or non-perturbative contributions
in the Wilson coefficients could conceivably be much
larger than anticipated. Presumably those would also lead
to larger lifetime differences among beauty hadrons
than anticipated. One would have to understand, however,
why these unorthodox effects are larger in $B$ than
in $\tau$ decays, rather than the other way around.
A less exotic possibility would be that the next-to-next-to-leading
perturbative terms in the Wilson coefficients are
considerably larger than expected on general grounds.
In principle this can be checked by a straightforward
analysis. Alas, in practice the necessary computations
appear to be rather forbidding.

(iii) `The exciting resolution':
New Physics controls 20\% of all $B$ decays! Obviously
one would expect that such a massive intervention of
new dynamics would lead to many signatures, like
charm content both in inclusive as well as exclusive
decays.

\vspace{1cm}
{\bf Acknowledgements:}
Useful and stimulating discussions with V. Braun,
M. Danilov, R. Poling,
N. Uraltsev and
M. Voloshin are acknowledged.
M. Voloshin informed us that explicit mechanisms for
explaining the lower than expected semileptonic
branching ratio for $B$ mesons in terms of New Physics are
under consideration at present (B. Smith, M. Voloshin, to be
published).
This work is supported in part by DOE
under the grant DOE-AC02-83ER40105 and by the NSF under the grant
PHY 92-13313.

\vspace{1cm}

{\bf Figure Captions}

\noindent Fig.1: Diagram where the average off-shellness is
below $m_b^2$.

\noindent Fig.2: Diagram with a typical off-shellness around $m_b^2$.

\noindent Fig.3a: Diagram for Weak Annihilation in $B^0$ decays;

\noindent Fig.3b: Diagram for Pauli Interference in $B^-$ decays.

\noindent Fig.4: Diagram generating the operator ${\cal O}_E$.

\end{document}